\documentclass[conference]{IEEEtran}
\IEEEoverridecommandlockouts
\usepackage{cite}
\usepackage{amsmath,amssymb,amsfonts}
\usepackage{algorithmic}
\usepackage{graphicx}
\usepackage{textcomp}
\usepackage{xcolor}
\usepackage{comment}
\def\BibTeX{{\rm B\kern-.05em{\sc i\kern-.025em b}\kern-.08em
    T\kern-.1667em\lower.7ex\hbox{E}\kern-.125emX}}

\usepackage{url}

\usepackage{etoolbox}
\usepackage{listings}
\usepackage{color}
\definecolor{codegray}{rgb}{0.5,0.5,0.5}
\definecolor{codepurple}{rgb}{0.58,0,0.82}
\definecolor{backcolour}{rgb}{0.95,0.95,0.92}

\lstdefinestyle{sql}{
    language=SQL,
    backgroundcolor=\color{backcolour},
    commentstyle=\color{codegray},
    keywordstyle=\color{blue},
    stringstyle=\color{codepurple},
    basicstyle=\ttfamily\footnotesize,
    breakatwhitespace=false,
    breaklines=true,
    captionpos=b,
    keepspaces=true,
    showspaces=false,
    showstringspaces=false,
    showtabs=false,
    tabsize=2
}

\lstdefinelanguage{yaml}{
  keywords={true,false,null,y,n},
  keywordstyle=\color{blue},
  basicstyle=\ttfamily\footnotesize,
  sensitive=false,
  comment=[l]{\#},
  morecomment=[s]{/*}{*/},
  commentstyle=\color{codegray},
  stringstyle=\color{codepurple},
  morestring=[b]',
  morestring=[b]",
  morestring=[s]{>}{\n},
  morestring=[s]{|}{\n}
}

\lstdefinestyle{yaml}{
    language=yaml,
    backgroundcolor=\color{backcolour},
    breakatwhitespace=false,
    breaklines=true,
    captionpos=b,
    keepspaces=true,
    showspaces=false,
    showstringspaces=false,
    showtabs=false,
    tabsize=2
}

\makeatletter
\newcommand{\linebreakand}{%
  \end{@IEEEauthorhalign}
  \hfill\mbox{}\par
  \mbox{}\hfill\begin{@IEEEauthorhalign}
}
\makeatother

\usepackage{fancyhdr}

\fancypagestyle{FirstPage}{
    \fancyhf{}%

    \fancyhead[L]{\parbox[b]{\textwidth}{\footnotesize Accepted to the 17th ACM/IEEE International Symposium on Empirical Software Engineering and Measurement (ESEM 2023), October 26--27, 2023, New Orleans, Louisiana, United States}}%
    \fancyfoot[L]{\parbox[t]{\textwidth}{\footnotesize \textcopyright{} 2023 IEEE. Personal use of this material is permitted. Permission from IEEE must be obtained for all other uses, in any current or future media, including reprinting/republishing this material for advertising or promotional purposes, creating new collective works, for resale or redistribution to servers or lists, or reuse of any copyrighted component of this work in other works.}}%
}

\begin{document}

\title{Evaluating the Impact of ChatGPT on Exercises of a Software Security Course}

\author{\IEEEauthorblockN{1\textsuperscript{st} Jingyue Li}
\IEEEauthorblockA{\textit{Dept. of Computer Science} \\
\textit{Norwegian Univ. of Science and Technology (NTNU)}\\
Trondheim, Norway\\
jingyue.li@ntnu.no}
\and
\IEEEauthorblockN{2\textsuperscript{nd} Per Håkon Meland}
\IEEEauthorblockA{\textit{Dept. of Computer Science} \\
\textit{NTNU}\\
Trondheim, Norway \\
per.hakon.meland@ntnu.no}
\and
\IEEEauthorblockN{3\textsuperscript{rd} Jakob Svennevik Notland}
\IEEEauthorblockA{\textit{Dept. of Computer Science} \\
\textit{NTNU}\\
Trondheim, Norway \\
jakob.notland@ntnu.no}
\linebreakand %
\IEEEauthorblockN{4\textsuperscript{th} André Storhaug}
\IEEEauthorblockA{\textit{Dept. of Computer Science} \\
\textit{NTNU}\\
Trondheim, Norway \\
andre.storhaug@ntnu.no}
\and
\IEEEauthorblockN{5\textsuperscript{th} Jostein Hjortland Tysse}
\IEEEauthorblockA{\textit{Dept. of Computer Science} \\
\textit{NTNU}\\
Trondheim, Norway \\
jostein.h.tysse@ntnu.no}

}

\maketitle
\thispagestyle{FirstPage} %

\begin{abstract}
Along with the development of large language models (LLMs), e.g., ChatGPT, many existing approaches and tools for software security are changing. It is, therefore, essential to understand how security-aware these models are and how these models impact software security practices and education. In exercises of a software security course at our university, we ask students to identify and fix vulnerabilities we insert in a web application using state-of-the-art tools. After ChatGPT, especially the GPT-4 version of the model, we want to know how the students can possibly use ChatGPT to complete the exercise tasks. We input the vulnerable code to ChatGPT and measure its accuracy in vulnerability identification and fixing. In addition, we investigated whether ChatGPT can provide a proper source of information to support its outputs. Results show that ChatGPT can identify 20 of the 28 vulnerabilities we inserted in the web application in a white-box setting, reported three false positives, and found four extra vulnerabilities beyond the ones we inserted. ChatGPT makes nine satisfactory penetration testing and fixing recommendations for the ten vulnerabilities we want students to fix and can often point to related sources of information.  
\end{abstract}


\section{Introduction}
Due to the risk and cost caused by software vulnerabilities, many universities have started educating software developers to develop secure code in the first place. The mandatory software security course for all IT students at our university teaches students how to incorporate security in each software development phase. Currently, the course focuses on security issues related to web applications. One critical module of the course is to teach students to understand the vulnerabilities listed in the open web application security project (OWASP) top 10 \cite{OWASP} and to identify and fix the vulnerabilities using state-of-the-art tools. 

In the current course exercises, we teach students to use several tools to identify and fix vulnerabilities inserted in a web application. Students often use, e.g.,  Firefox web developer tool \cite{FirefoxTool},  Postman~\cite{postman}, and Zap \cite{ZAP}, to gather web application information. Students also use scanners, e.g., Nessus \cite{Nessus}, to scan the web application to identify vulnerabilities. In addition, students read the OWASP web security testing guide (WSTG) \cite{wstg} to learn how to black-box test the web application and review the code to identify the vulnerabilities. 

After ChatGPT \cite{ChatGPT}, especially the GPT-4 model \cite{GPT4}, is available on 14 March 2023, we suspect that students may use it to complete the exercise tasks without needing to learn other tools or read the testing guide \cite{wstg}. Becker et al. \cite{becker2022programming} identified issues and opportunities in using AI-powered tools to teach an introductory programming course. Results of \cite{10.1145/3511861.3511863} showed that AI-powered tools could outperform students in answering exam questions. Tony et al. \cite{tony2022conversational} found that chatbots could help identify vulnerabilities in the code, but with limitations. Studies, e.g., \cite{chen2021evaluating} and \cite{fischer2017stack}, showed that many of the codes generated from chatbots were vulnerable. However, to our knowledge, no study has investigated the impact of ChatGPT on teaching software vulnerability identification and fixing in a university course setup. In the coming years, we assume many students can access ChatGPT \cite{ChatGPT} or similar tools. Thus, we are motivated to understand ChatGPT's impact on teaching software security, especially its impact on the exercise design.  For now, we focus on GPT-4 \cite{GPT4}  because it allows for a context window of 8k tokens, meaning students can copy and paste a large chunk of code into it and let it identify and fix the vulnerabilities in the code for them. Our research questions are:

\begin{itemize}
    \item RQ1: How accurate is ChatGPT in vulnerability identification? 
    \item RQ2: How accurate is ChatGPT in recommending penetration test cases? 
    \item RQ3: How accurate is ChatGPT in  recommending vulnerability fixes? 
    \item RQ4: How often can ChatGPT provide sufficient sources of information related to software vulnerability identification and fixing?
\end{itemize}

To answer the research questions, we asked teaching assistants of the course to input the vulnerable code to GPT-4 and use its prompts to guide vulnerability identification and fixing. We then calculated the identification and fixing recommendation accuracy of ChatGPT and summarized the fruitfulness of the source of information it provided. 

The study results show that ChatGPT can provide satisfactory results to identify and fix the inserted vulnerabilities, although there are still rooms to improve. The contributions of the study are: 
\begin{itemize}
    \item We have identified the pros and cons of using ChatGPT to identify software vulnerabilities and propose fixing recommendations. 
    \item To our knowledge, it is the first study investing in ChatGPT's impact on teaching software security, which gives valuable insight into updating the design of similar courses in the LLM era. 
    \item Based on  the results of this study, we provide visions for future research in software security and IT education. 
\end{itemize}

The rest of the paper is organized as follows. Section \ref{related-work} shows related work. Section \ref{background} presents the background of the software security course. Section \ref{research-design} explains research design. The results are presented in Section \ref{reseach-results} and are discussed in Section \ref{discussion}. Section \ref{conclusions} concludes the study, and Section \ref{future-work} presents future work. 

\section{Related work}
\label{related-work}

An overview of the most recent studies relevant to our work is given in Table \ref{tab:relwork}.

\begin{table}[ht!]
    \centering
    \caption{Overview of related work}
    \begin{tabular}{p{0.4cm}|p{0.8cm}|p{3.6cm}|p{2.1cm}}
         \textbf{Ref.} & \textbf{Chatbot} & \textbf{Tests} & \textbf{Results}  \\
         \hline
            \cite{tony2022conversational} &  SKF & Performance, efficiency and user acceptance of chatbot vs. Internet search for a Java web application with XSS and SQL injection vulnerabilities. & Internet search outperformed SKF. \\ 

            \cite{10.1145/3511861.3511863} &  OpenAI Codex & Chatbot vs. students in solving programming questions. & Codex outperformed students. \\ 

            \cite{chen2021evaluating} &  OpenAI Codex &  Comparison of Python code writing capabilities between Codex, GPT-3, and GPT-J. & Codex outperformed the others. \\ 

            \cite{perry2022users} &  OpenAI Codex & Participants with chatbot support given programming tasks in Python, JavaScript, and C. & Participants wrote code with more vulnerabilities using chatbot than not using it. \\

    \end{tabular}
    \label{tab:relwork}
\end{table}

Tony et al. \cite{tony2022conversational}  analyzed the effectiveness of the security knowledge framework (SKF) chatbot stemming from OWASP to find and fix security vulnerabilities by giving participants source code with known vulnerabilities. However, they only focused on Cross-Site Scripting (XSS) and SQL injection vulnerabilities. Their results showed that only three out of 15 participants arrived at the right fix with chatbot support, while seven out of 15 participants solved the task using a (manual) Internet search.

Finnie-Ansley et al.~\cite{10.1145/3511861.3511863} showed that OpenAI's Codex outperformed 80\% of students in exam questions related to the introductory programming course, which could provide low-risk/high rewards for students focusing on getting good grades rather than developing an understanding, leading to academic misconduct.

Chen et al. \cite{chen2021evaluating} analyzed the Python code-writing capabilities of OpenAI's Codex and showed that OpenAI's Codex could produce vulnerable or misaligned code. Santhanam et al. \cite{santhanam2022bots} found that around 2/3 of the chatbots for code generation used Stack Overflow as their main information source. Fischer et al. \cite{fischer2017stack} figured out that out of 1.3 million Android applications using code snippets from Stack Overflow, 97.9\% contained at least one insecure code snippet.  

Perry et al. \cite{perry2022users} examined how developers chose to
interact with AI code assistants and how
those interactions caused security mistakes. They found that their participants with chatbot access wrote less secure code than those without such access, despite that the participants themselves thought it would be vice-versa. This showed that such tools could give users a false sense of the security of the code. %

\section{Course background}
\label{background}

Although the detailed requirements and grading criteria of the mandatory exercises of the software security course vary each year, the exercises are usually designed as follows. 
\begin{itemize}
    \item We develop a new web application and insert around 20 to 30 vulnerabilities listed in the OWASP top 10 \cite{OWASP} each year. The web application in the Spring 2023 semester simulates functionality for refugees to find volunteer help services. Certified volunteers register their wishes and skills to provide help services through the web application. Administrators of the web application handle skill certifications to verify the volunteers. The web application uses Django 4.0.8 with Django Rest Framework 3.13.1 as the back end. The front-end code is developed using React 4.0.3. To route the requests to the front end and back end, we use NGINX 1.23.1 as a reverse proxy. The web application is hosted at the internal Gitlab repository of our university and is only accessible with students' university credentials. Thus, we believe the vulnerable web application used in 2023 is not included in ChatGPT's training dataset. 
    \item In the first module of the exercise, students must first use tools to collect the web application's information and possibly make a page map. Based on the page map, students must identify the inserted vulnerabilities using code reviews and black-box testing approaches and tools. Students need to write reports to describe how the tests are performed to identify the vulnerabilities and the location of the vulnerable codes. Besides reporting the vulnerabilities and their locations, students must also provide vulnerabilities' WSTG \cite{wstg} code to refer the vulnerabilities to proper categories.  
    \item In the exercise's second module, students must fix the identified vulnerabilities. To reduce the students' effort to fix the vulnerabilities, we do not ask them to fix all vulnerabilities. We make a list of, for example, ten vulnerabilities and let students fix only those. By limiting and predefining the set of vulnerabilities to identify and fix, we make it easier to make a solutions guide and coach our teaching assistants.
\end{itemize}

As more than 200 students take this course each year, the exercises are often organized in groups containing one to three students. We give grades to students based on reading their reports and counting the number of vulnerabilities they identify and fix correctly. We do not punish students if they report false positive vulnerability identification results. Sometimes, students find and report more vulnerabilities than we have inserted.
If students identify extra vulnerabilities, we give them bonus points up to the full score of the exercises. In parallel with these exercises, we also offer students free access to a software game, Secure Code Warrior \cite{SourceCodeWarrior}, which teaches software security, as a supplement. An example feature of the Secure Code Warrior game is that it provides multiple-choice questions and asks students to identify a location of code that is vulnerable among multiple code locations.

In previous years, the grades of the exercises were added to the course's final grades. From the Spring 2023 semester, the exercises grade are only used to qualify a student to take the final written exam if the student's exercise grade passes a threshold. The final written exam counts as 100\% of the course grade.

\section{research design}
\label{research-design}
This section explains the detailed design to answer each research question. In general, we designed the prompts for ChatGPT in the same way that the exercises are structured and how we expect students to approach them. We standardized the prompts such that we first asked our question before pasting the relevant code. We asked follow-up questions until we could verify whether ChatGPT could provide a satisfactory answer.

\subsection{Design and implementation to answer RQ1}\label{RQ1-design}

To answer RQ1, one teaching assistant, who had inserted the vulnerability in the web application as the preparation of the exercises, evaluated how accurately ChatGPT identified vulnerabilities in the white-box settings. The teaching assistant first copied and pasted the code of each front- and back-end file into ChapGPT and asked it to identify the vulnerabilities in the code. As each file contains fewer than 8k tokens, the teaching assistant did not need to split the codes in the file when inputting them into ChatGPT. An example prompt the teaching assistant used was as follows. 

\textit{This is the settings.py for a Django project. Can you find any vulnerabilities and list them according to OWASP top 10 or with the OWASP WSTG id? Also tell me explicitly which line of code is the problem.}

In addition, the teaching assistant asked ChatGPT to provide the WSTG \cite{wstg} code as follows. 

\textit{Do you have the owasp wstg code for these vulnerabilities?}

Sometimes, ChatGPT did not give a complete list of vulnerabilities in the first place. So, the teaching assistant asked an additional question as follows. 

\textit{Do you find any other vulnerabilities in the file?}
\textit{Do you find anything else?}

As students often use the WSTG \cite{wstg} as a checklist to identify vulnerabilities, the teaching assistant also simulated such a process and asked more concrete questions as follows to identify a particular type of vulnerability in the code:

\textit{How about the token lifetimes, are they sufficient?}

After applying similar prompts on each file, the teaching assistant calculated the accuracy of the ChatGPT's vulnerability identification.  

\subsection{Design and implementation to answer RQ2}\label{RQ2-design}
As aforementioned, we specified only ten vulnerabilities for students to fix. To answer RQ2, we decided to focus on asking ChatGPT to propose penetration test cases only on those ten vulnerabilities to limit the teaching assistant's effort in performing this study. Another teaching assistant, who assisted in the vulnerability insertion, asked ChatGPT to propose penetration test case to answer RQ2. The example prompts for penetration testing are as follows:

\textit{How can I exploit the following code by XSS?}

\textit{Show me an example html injection.}

\textit{How can I observe unencrypted HTTP network traffic from my computer?}

\subsection{Design and implementation to answer RQ3}\label{RQ3-design}

The teaching assistant, who answered RQ2, also performed the study to answer RQ3. The example prompts are as follows to let ChatGPT first identify the vulnerability through penetration testing and then propose fixes.

\textit{Does my login function below have any lockout functionality?} 

Then, the teaching assistant asked: 

\textit{How can I test whether my application has lockout functionality in a black-box fashion?} 

At last, the teaching assistant asked:

\textit{Show me how to use Django-axes to solve this problem.}

For each of the ten vulnerabilities, the teaching assistant used similar prompts and counted the accuracy of the proposed penetration test cases by ChatGPT to answer RQ2 and fixing solutions proposed by ChatGPT to answer RQ3.

\subsection{Design and implementation to answer RQ4}\label{RQ4-design}

To answer RQ4, a third teaching assistant first asked ChatGPT to identify security vulnerabilities in each file using prompts as follows.

\textit{I want you to identify security vulnerabilities in the following file plus the code with vulnerabilities}.  

The ChatGPT usually supplies a response with the results. The teaching assistant then asked:

\textit{Can you provide sources for that information?} 

The teaching assistant did not include any notion of Django, OWASP, etc. This is to ensure the results are not biased towards any particular source. After asking ChatGPT to identify security vulnerabilities in two front-end files, three back-end files, one Django setting file, and one NGINX configuration file, the teaching assistant counted the number of cases in which ChatGPT provided sufficient source of information.

\section{Research results}
\label{reseach-results}

\subsection{Results of RQ1}\label{RQ1-result}

Results of vulnerability identification in the white-box settings showed that GPT-4 found 16 of our 28 listed vulnerabilities on the first try. With some additional more specific prompts, four more were found. 
The ones identified by ChatGPT include, for example, sensitive information sent via unencrypted channels, default admin password, and unlimited login attempts without a lockout. A complete list of the identified vulnerabilities is in the file \footnote{\label{note1} \url{https://doi.org/10.6084/m9.figshare.23576370}}.

The eight vulnerabilities that were not found included, for example, unsafe user registration process (multiple users can have the same email), username enumeration with error messages from user registration, and reset password does not validate token correctly. A complete list of the missed vulnerabilities is also in the file \footref{note1}. A possible explanation of the missed vulnerabilities could be that ChatGPT did not fully understand the code's behavior or did not access every line of the code in one prompt. The built-in Django methods and imported codes caused several of these vulnerabilities. However, in some cases, as with ``reset password does not validate'', ChatGPT totally missed a logical error where the random token was not used to validate the reset password link.

ChatGPT reported three false positives, namely, improper error handling, insecure direct object reference in the document download method, and broken access control – cross-origin resource sharing (CORS) settings. The allowed hosts' CORS settings were reported as a vulnerability by ChatGPT. However, in our configuration to run the application on the server, it is not a vulnerability. If students blindly trust the results from ChatGPT, false positives may confuse them since they cannot perform the exploits for the report. 

Beyond the vulnerabilities we inserted, ChatGPT identified four extra ones, namely, admins able to see more information than they should, no CAPTCHA for registration, SQLite used in production, and improper exception handling – broad exception catching. 

To get a full score on the exercise, we required the students
to find 15 vulnerabilities. The results of RQ1 mean that ChatGPT can pass the threshold.

\subsection{Results of RQ2}\label{RQ2-result}

ChatGPT proposed satisfactory methods to penetration test the vulnerabilities in nine of the ten cases. Data in the file \footnote{\label{note2} \url{https://doi.org/10.6084/m9.figshare.23628879}} shows an overview of the tests. ChatGPT struggled with answering our questions when:

\begin{enumerate}
    \item It has ethical issues showing penetration test cases.
    \item The number of files and code length exceeds the input limit. 
    \item ChatGPT might assume that our application has some default components. On the other hand, it overlooked details in our code.
\end{enumerate}

The combination of the issues caused the flawed answer from ChatGPT. The flawed penetration test is about SQL injection. ChatGPT proposed the following:

\begin{lstlisting}[style=sql]
request_id = "1' UNION SELECT * FROM auth_user WHERE username = 'admin' AND '1' = '1"
\end{lstlisting}

This is an injection, and it will execute in the database. However, there is an issue that the auth\_user table is not part of our models. We have extended Djangos default models and named the table users\_user instead. Therefore the injection will cause an error, and it does not extract any useful information or does any harm to the database. When we provide ChatGPT with more context, it will stop providing injections because of ethical issues:

\textit{Please note that providing working SQL injection examples might encourage malicious activities, which is against OpenAI's policy to promote such behavior.}

\subsection{Results of RQ3}\label{RQ3-result}

As shown in the file \footref{note2}, ChatGPT provided satisfactory recommendations for fixing the vulnerabilities in nine of the ten cases. 
Regarding the flawed mitigation recommendation of ``sensitive information sent over unencrypted channels'', ChatGPT struggled to provide a complete answer. We provided seven files relevant to configuring HTTPS:

\begin{itemize}
    \item .env: provides PORT\_PREFIX, GROUP\_ID, DOMAIN, PROTOCOL (PORT\_PREFIX and GROUP\_ID compose the server port).
    \item Dockerfile (front end): Composes the .env variables into an URL to send API requests.
    \item Dockerfile (back end): Passes the .env variables to Django.
    \item settings.py: composes the .env variables for user redirects.
    \item Dockerfile (nginx): Passes nginx.conf.
    \item nginx.conf: Configures the reverse proxy.
    \item docker-compose.yml: Builds the project using the Dockerfiles and defines port forwarding.
\end{itemize}

ChatGPT suggested a viable nginx.conf and showed how to generate the certificate/key. However, it suggested a flawed port configuration as below for docker-compose.yml, where two ports were forwarded to the same port, causing an error:

\begin{lstlisting}[style=sql]
ports:
  - ${PORT_PREFIX}${GROUP_ID}:80
  - ${PORT_PREFIX}${GROUP_ID}:443
\end{lstlisting}

By adjusting the nginx.conf file to incorporate the HTTPS configuration suggested by ChatGPT and simultaneously limiting port forwarding to only port 443, we caused CORS errors to emerge from Django. The root cause was that ChatGPT did not acknowledge the critical role of the PROTOCOL variable in the .env file. This variable delineates whether the URL would employ HTTP or HTTPS when making API calls and functions as a mechanism to establish the list of allowed origins in Django. When ChatGPT recommended the implementation of HTTPS via redirection of HTTP requests, as opposed to direct utilization of HTTPS, it created a conflict between the incoming request and the expected allowed origin.

The new HTTPS endpoint was intended to replace the previously forwarded HTTP endpoint using the same port. By guiding the students to modify the PROTOCOL to HTTPS and use a pure HTTPS endpoint, we had already circumvented the issues introduced by ChatGPT. This instance exemplified ChatGPT's challenges in fully grasping the context and its intricate interdependencies.

\subsection{Results of RQ4}\label{RQ4-result}

The ChatGPT provided sufficient sources of information in all cases and pointed to different resources for different vulnerabilities as follows.
\begin{itemize}
    \item For vulnerabilities in the settings file, ChatGPT provided references to the Django documentation and some GitHub repositories.
    \item For back-end-related vulnerabilities, ChatGPT provided mostly references to the Django documentation and OWASP top 10 \cite{OWASP}.
    \item For front-end-related vulnerabilities, ChatGPT provided mostly references to OWASP top 10 \cite{OWASP} and MDN (Mozilla Developer Network).
    \item For NGINX-related vulnerabilities, ChatGPT provided mostly references to MDN (Mozilla Developer Network) and the NGINX documentation.
\end{itemize}

In summary, ChatGPT could provide relatively good sources. All the suggested sources were relevant websites. For example, it referenced the Django security documentation and OWASP top 10 \cite{OWASP}. On occasions, it even provided Stack Overflow references. However, some of the sources are a bit outdated. For example, the OWASP top 10 is from 2017.

However, when the teaching assistant asked:

\textit{Where did you get that information from?} 

The answers from ChatGPT were more generic, such as:

\textit{As an AI language model, my knowledge is based on the vast amount of text data that I have been trained on, which includes information from a variety of sources such as websites, books, articles, and other documents. Specifically, for Django-related information, I rely on the official Django documentation and various articles, tutorials, and blog posts written by experienced Django developers and experts. }

\textit{My training data includes text sources up to September 2021. It is important to note that my knowledge does not include the latest updates or developments that may have occurred after this date. For the most up-to-date information, it's always a good idea to consult the official Django documentation or other trusted sources.}

\section{Discussion}
\label{discussion}

\subsection{Comparison with related work}

Compared to the study \cite{tony2022conversational} using the SKF chatbot for security identification, our study covers many more types of vulnerabilities than Cross-Site Scripting and SQL injection. Our study shows that ChatGPT can precisely identify many types of vulnerabilities in large file chunks but is still limited in identifying cross-file vulnerabilities across the codebase.   

Regarding \cite{becker2022programming} focusing on the impact of AI-based approach on software engineering education, our results raise a similar alarm, i.e., course exercise design and grading shall be updated in the LLM era. However, our study investigates the issue from identifying and fixing software vulnerability perspectives rather than automatically generating and completing functional code. 

The results of \cite{chen2021evaluating, perry2022users} showed that the code generated from
LLMs might be vulnerable. We applied ChatGPT to identify
vulnerabilities and found that ChatGPT could miss identifying
a few types of vulnerabilities, indicating that vulnerable code
in ChatGPT’s training dataset might also degrade its capability
to detect vulnerabilities.

\subsection{Implication}

Results of RQ1 indicate ChatGPT excels at detecting vulnerabilities within its context window, aligning with OpenAI's findings~\cite{openai2023gpt4}. However, detecting complex vulnerabilities across multiple files can be challenging due to its 8k token limit, despite being double that of similar models. This limit restricts the codebase analyzed, requiring knowledge of which files to assess together for cross-file vulnerabilities. Access to the 32k version of GPT-4 could enhance detection capabilities by allowing larger codebase analysis, thereby streamlining the detection process.

Results of RQ2 and RQ3 show ethical concerns arising from penetration testing questions. Simultaneously, the input limit affects the context we can provide for ChatGPT, potentially leading to false assumptions or overlooked code characteristics. Results of RQ4 demonstrate ChatGPT's ability to provide students with adequate information to support their solutions.

This study suggests that the introduction of ChatGPT presents teaching challenges in software security. If we continue grading based on counted vulnerabilities found, exploited, and fixed, students may score well by leveraging ChatGPT. We could consider blocking access to ChatGPT during exercises, but this could prove difficult to enforce, given that students complete the exerciseses mainly from home.

We could revise our exercise grading system to include report writing, where students compare various tools and solutions. However, this would disrupt our existing quantitative approach, requiring new procedures to ensure fair and consistent evaluations by the teaching assistants. This is crucial, especially if exercise grades impact course grades directly. Incorporating qualitative assessments, like discussions, would require a framework to maintain fairness and consistency. Moreover, the possibility of students leveraging AI to answer discussion questions should also be considered.

Our study reveals ChatGPT's difficulty in detecting cross-file vulnerabilities in codebases. While incorporating more such vulnerabilities could increase the challenge of using chatbots for tasks, this may make some vulnerabilities less realistic, as real ones often are contained within a single file. Should we require students to locate cross-file vulnerabilities, we may need to frame the exercise questions to hint at their location.

ChatGPT can enhance students' comprehension of software security issues. We currently introduce penetration testing tools and automated scanners, stressing their limitations as per OWASP's testing guide~\cite{owasp2014testingGuide}. We can treat ChatGPT similarly and teach students its use alongside its limitations.

Today, students must find 15 of 28 inserted vulnerabilities to get a passing grade for the exercise due to the limited time they can spend on the course. In the future, by teaching students to use ChatGPT to identify and fix vulnerabilities, we can raise the threshold to require students to find more vulnerabilities. Thus, we can motivate them to understand more types of vulnerabilities.

\subsection{Threat to validity}
The teaching assistants and course teachers first agreed upon the study design to avoid data analysis biases. The data analysis results were shared among the teaching assistants using the functions of \url{https://sharegpt.com/} and were cross-checked. Although the study focused on only one software security course and one web application, we believe the insights from this study can be generalized to other software security courses focusing on web application security. The web application we focused on includes typical features, such as registration, login, password encryption, session management, etc., related to web application security and popular OWASP top 10 vulnerabilities.

\section{Conclusions}
\label{conclusions}
This study piloted ChatGPT to identify its possible impacts on teaching a software security course, especially on the exercises of the course. The results show that students can easily get good exercise grades by simply asking ChatGPT and copying the results to the reports. The results also provide valuable insights into the advantages and limitations of using ChatGPT for identifying and fixing typical web application security issues.

\section{Future work}
\label{future-work}
There is a wide range of research tasks that can be envisioned 
in the field of chatbot development and secure coding. 
In the coming year, we plan to provide a selected group of students access to ChatGPT, allow them to do exercises using it, and then interview them to identify the pros and cons of ChatGPT from students' perspectives. We encourage the community to perform similar studies to establish more empirical evidence, for instance, identifying local variations, performance for different programming languages and trends over time.
We also foresee studies where we can evaluate and compare large language models from different organizations as they continue to become available. A challenge here is that the technology and datasets develop so rapidly that results from most benchmark experiments are quickly outdated.
A framework for structuring the research and maybe even automating experiments would be useful assets to the research community.

\section*{Data availability}
\bgroup
\appto\UrlNoBreaks{\do\/\do\:}%
The source code of the web application is available at \url{https://doi.org/10.6084/m9.figshare.23576364}. All the prompts and ChatGPT results related to this study are available at \url{https://doi.org/10.6084/m9.figshare.23629455}.
\egroup

\section*{Acknowledgment}
The research conducted in this paper has partly been related to the CyberSecPro project under the European Union’s Digital Europe Programme (DEP) (grant agreement No 101083594).

\bibliographystyle{IEEEtran}
\bibliography{references.bib}

\end{document}